\documentclass[12pt,a4paper]{article}
\newcommand\be{\begin{equation}}
\newcommand\ee{\end{equation}}
\newcommand\bea{\begin{eqnarray}}
\newcommand\eea{\end{eqnarray}}
\newcommand\bei{\begin{itemize}}
\newcommand\ei{\end{itemize}}
\newcommand\ket[1]{|#1\rangle}
\newcommand\bra[1]{\langle #1|}

\begin{document}
\title{Quantum Time Evolution in Terms of Nonredundant Expectation Values}
\author{Stefan Weigert\\
Institut de Physique, Universit\'e de Neuch\^atel\\
Rue A.-L. Breguet 1, CH-2000 Neuch\^atel, Switzerland\\
\texttt{stefan.weigert@iph.unine.ch}}
\date{March 1999}
\maketitle
\begin{abstract}
Each scheme of state reconstruction comes down to parametrize the state of a quantum system by expectation values or probabilities directly measurable in an experiment. It is argued that the time evolution of these quantities provides an unambiguous description of the quantal dynamics. This is shown explicitly for a single spin $s$, using a quorum of expectation values which contains no redundant information. The quantum mechanical time evolution of the system is rephrased in terms of a closed set of linear first-order differential equations coupling $(2s+1)^2$ expectation values. This `realization' of the dynamical law refers neither to the wavefunction of the system nor to its statistical operator. 
\end{abstract}
For a quantum system with statistical operator $\hat \rho$, it is straightforward to determine the expectation value $\langle\,  {\widehat {\cal A}} \, \rangle$ of an operator 
${\widehat {\cal A}} $ according to
\be
\langle \, {\widehat {\cal A}} \, \rangle_\rho 
 = \mbox{ Tr }\left[ {\widehat {\cal A}} \, \hat \rho \right] \, .
\label{expval}
\ee
Methods of {\em state reconstruction} \cite{bertrand+87,risken+89,leonhardt97,raymer97} solve the inverse problem: the unknown state $\hat \rho$ of the quantum system is expressed as a function of the expectation values of properly chosen observables ${\widehat {\cal Q}}^{\, j}$, which constitute a {\em quorum} $\cal Q$. The resulting equivalences      
\be
\hat \rho \, \Leftrightarrow \,  \{   \langle {\widehat{\cal Q}}^{\,j} \rangle,  j \in I \} \, ,
\qquad {\widehat{\cal Q}}^{\, j} \in {\cal Q} \, , 
\label{suggestive}
\ee
with $j$ taking on values from a discrete or continuous set $I$ of labels, are more than 
just mathematical beauties---they have been used in the laboratory to reconstruct correctly states of various quantum systems \cite{kurtsiefer+97,dunn+95,smithey+93}. For example, the state of an ion in a Paul trap has been identified \cite{leibfried+96} by a method realizing (\ref{suggestive}) on the basis of Wigner functions.    

The purpose of the present paper is to point out that a parametrization as in (\ref{suggestive}) suggests a conceptually interesting way to describe the time evolution of a quantum system {\em without} invoking its density matrix or wave function. Instead, only  directly measurable quantities, that is, expectation values of hermitean operators are involved. The argument will be given in general terms first, specifying neither the system at hand nor a particular method of state reconstruction. In the main part of the paper the example of a single spin $s$ is worked out explicitly, followed by a discussion putting the results into perspective.

The `quantum mechanical Liouville' equation \cite{sakurai85},
\be
\frac{d \hat \rho }{dt}   = - \frac{i}{\hbar} [ \widehat{H} \, , \, \hat \rho \, ] \, ,
\label{schroe}
\ee
describes the time evolution of a quantum system with Hamiltonian operator $\widehat{H}$ and statistical operator $\hat \rho$. The state ${\hat \rho}_0 $ at time $t_0$ is transported to ${\hat \rho}_1$ at time $t_1$ along a smooth path in state space. Suppose that the operators $\{ {\widehat{\cal Q}}^{\, j} \}$ provide a quorum ${\cal Q}$ for all possible states of the system at hand. Then, each state ${\hat \rho}_t$ on the path between ${\hat \rho}_0 $ and ${\hat \rho}_1 $ is characterized uniquely by the set of expectation values  $\{ \langle {\widehat{\cal Q}}^j \rangle_t \}$. In other words, the path ${\hat \rho}_t$ in state space has an unambiguous image in the {\em space of expectations} 
$\{ \langle {\widehat{\cal Q}}^{\, j} \rangle \}$. This path is expected to arise as the solution of a dynamical law in this space \cite{weigert98}:  
\be
\frac{d}{dt} \langle {\widehat{\cal Q}}^{\, j} \rangle_t 
= {\cal D}_{\widehat{H}}^{\, j} 
      \left( \{ \langle {\widehat{\cal Q}}^{\, j} \rangle_t  \}  \right) \, ,
\label{newdynamics}
\ee
where the function ${\cal D}$ will depend on both the  Hamiltonian $\widehat{H}$ of the system and the quorum $\cal Q$. Subsequently, time dependent expectation values of {\em arbitrary} operators ${\widehat {\cal A}}$ can be expressed in terms of the 
$\{ \langle {\cal {\widehat Q}}^{\, j} \rangle_t \}$ simply by using Eq.\ (\ref{suggestive}) to  eliminate ${\hat \rho}_t$ in favor of the elements of the quorum.  

In the following, an explicit form of Eq.\ (\ref{newdynamics}) will be derived for a spin $s$ using a nonredundant quorum. Quantum mechanically, the spin is described by a vector operator $\widehat {\bf S} 
\equiv \hbar \widehat{\bf s}$, the components of which satisfy the commutation relations of the algebra $su(2)$: $[ {\hat s}_x, {\hat s}_y ]= i{\hat s}_z, \ldots$ These operators act irreducibly in a Hilbert space $\mathcal{H}_s$ of (complex) dimension $(2s+1)$. The standard basis of the space $\mathcal{H}_s$ is given by the eigenvectors of the $z$ component  of the spin, ${\widehat S}_z = {\bf n}_z \cdot \widehat {\bf S}$, and they are denoted by $\ket{\mu, {\bf n}_z},$ $-s \leq \mu \leq s$ \cite{phases}. 

Observables are represented by hermitian operators, ${\widehat {\cal A}}^\dagger = {\widehat {\cal A}}$, all of which are linear combinations of polynomials in the operators ${\hat s}_x$, ${\hat s}_y$ and ${\hat s}_z$ of degree $2s$ at most. The ensemble of all hermitean operators acting on $\mathcal{H}_s$ can be considered as 
a vector space ${\cal A}_s$ of dimension $N_s = (2s+1)^2$. One well-known basis \cite{fano+59} for these operators is given by the {\em multipoles} ${\widehat K}_{lm}$, $ 0\leq l \leq 2s, \, -l\leq m \leq l$. They are associated with the group $SU(2)$: as a tensorial set, they transform in a simple way under rotations. In terms of those,
a hermitean operator ${\widehat {\cal A}}$ can be written as  
\begin{equation}
{\widehat {\cal A}} = \frac{1}{2s+1} \sum_{lm} A^{lm} {\widehat K}_{lm} \, ,
\label{expandO}
\end{equation}
with a unique set of expansion coefficients $A^{lm}$.

Given the multipoles ${\widehat K}_{lm}$, a second set of $N_s$ operators ${\widehat {\sf K}}^{lm}$ is known to exist such that 
\begin{equation}
\frac{1}{2s+1} \mbox{Tr}\left[  {\widehat K}_{lm} {\widehat {\sf K}}^{l'm'} \right] 
             = \delta_l^{l'} \delta_m^{m'} \, ,
\label{orthogonality}
\end{equation}
where the trace taken in the Hilbert space $\mathcal{H}_s$ defines the scalar product of two operators. The operators $\{ {\widehat {\sf K}}^{lm} \}$ constitute a second, {\em dual} basis of the space ${\cal A}_s$, being determined unambiguously \cite{greub63} by the original basis. In the case of multipoles, the elements of the dual basis are known explicitly: ${\widehat {\sf K}}^{lm} = {\widehat K}_{lm}^{\, \dagger}$. Using (\ref{orthogonality}), one easily determines the coefficients $A^{lm}$ in the expansion (\ref{expandO}):
\be
A^{lm} 
   = \mbox{ Tr } \left[ {\widehat {\cal A}} \, {\widehat {\sf K}}^{lm} \right] \, .
\label{coeffO}
\ee
There is a second expansion for selfadjoint operators ${\widehat {\cal A}}$ in terms of the dual basis. In analogy to Eqs. (\ref{expandO}) and (\ref{coeffO}) one can write any operator 
${\widehat {\cal A}}$ as a linear combination of the operators ${\widehat {\sf K}}^{lm}$ with coefficients ${\sf A}_{lm} = \mbox{ Tr } [ {\widehat {\cal A}} \, {\widehat K}_{lm} ]$ divided by $(2s+1)$.

Here the purpose is to express the dynamical evolution of a quantum spin in 
terms of expectation values. To this end, a basis different from the multipoles will be used. Denote the eigenstate of the operator ${\bf n} \cdot \hat {\bf s}$ along the direction  ${\bf n} = ( \sin \vartheta \cos \varphi, \sin \vartheta  \sin \varphi, \cos \vartheta)$ and with eigenvalue $s$ by 
\be
\ket{{\bf n}} 
        \equiv  \exp [ -i \, \vartheta \, {\bf m}(\varphi) \cdot {\hat {\bf s}} \, ] \, \ket{s,{\bf n}_z} \, ,
\label{defineaxes}
\ee
where ${\bf m}(\varphi) = (- \sin \varphi,\cos\varphi,0)$: the state $\ket{{\bf n} }$ is obtained from rotating the state $\ket{s,{\bf n}_z}$ about the axis ${\bf m}(\varphi)$ in the $xy$ plane by an angle $\vartheta$. In this way, a {\em coherent state} $\ket{ {\bf n}}$ is associated to each  point of the surface of the unit sphere \cite{arecchi+72}. The ensemble of all coherent states provides an overcomplete basis of the Hilbert space $\mathcal{H}_s$.

 The density matrix ${\hat \rho}$ of a spin $s$ is determined unambiguously if one performs appropriate measurements with a traditional Stern-Gerlach apparatus. Distribute $N_s = (2s+1)^2$ axes ${\bf n}_{\mu\nu}, -s \leq \mu,\nu \leq s$, over $(2s+1)$ cones about the $z$ axis with different opening angles such that the set of the $(2s+1)$ directions on each cone is invariant under a rotation about $z$ by an angle $2\pi/(2s+1)$. As shown in \cite{amiet+99/1}, an unnormalized statistical operator ${\hat \rho}$ is then fixed by measuring the $(2s+1)^2 $ relative frequencies $p_s({\bf n}_{\mu\nu}) = \bra{ {\bf n}_{\mu\nu}} \hat \rho \ket{ {\bf n}_{\mu\nu}}$, that is, by the expectation values of the statistical operator $\hat \rho$ in the coherent states $\ket{{\bf n}_{\mu\nu}}$. In other words, the projection operators 
\be
{\widehat Q}_{{\mu\nu}} 
        = \ket{{\bf n}_{{\mu\nu}}} \bra{{\bf n}_{{\mu\nu}}}   \, , 
          \qquad  -s \leq \mu,\nu \leq s  \, ,
\label{mixedrho}
\ee
constitute indeed a quorum ${\cal Q}$ for a spin $s$. This fact, when used for state reconstruction, 
defines an {\em optimal} method since exactly $(2s+1)^2$ numbers have to be determined experimentally which equals the number of free (real) parameters of the (unnormalized) hermitean density matrix $\hat \rho$. For the following, it is convenient to replace the labels $\mu$ and $\nu$ by a single index $n= (\mu\nu)$, say, with $1 \leq n \leq N_s$.  

The set of all unnormalized hermitean density matrices for a spin $s$ is just the set of all hermitean operators acting on the Hilbert space $\mathcal{H}_s$. Therefore, the quorum ${\cal Q}$ automatically provides a basis for hermitean operators $\widehat{\cal A}$:
\begin{equation}
{\widehat {\cal A}} = \frac{1}{2s+1} \sum_{n=1}^{N_s} A^{n} {\widehat Q}_{n} \, ,
\qquad 
A^{n}  = \mbox{ Tr } \left[ {\widehat {\cal A}} \, {\widehat {\sf Q}}^{n} \right] \, ,
\label{expandquorum}
\end{equation}
and the expansion coefficients $A^{n}$ involve operators ${\widehat {\sf Q}}^{n}$ {\em dual} to the elements of the original basis. In analogy to the example involving multipoles, there are orthogonality relations,
\begin{equation}
\frac{1}{2s+1} \mbox{ Tr }\left[  {\widehat Q}_{n} {\widehat {\sf Q}}^{n'} \right] 
= \delta_{n}^{n'}  \, , 
\qquad 1 \leq n,n' \leq N_s \, ,
\label{orthogonalityquorum}
\end{equation}
and a second expansion for hermitean operators is available:
\begin{equation}
{\widehat {\cal A}} = \frac{1}{2s+1} \sum_{n=1}^{N_s} {\sf A}_{n} {\widehat {\sf Q}}^{n} \, ,
\qquad 
{\sf A}_{n}  = \mbox{ Tr } \left[ {\widehat {\cal A}} \, {\widehat Q}_{n} \right] \, .
\label{expandquorumdual}
\end{equation}

Let us now consider the properties of the statistical operator $\hat \rho$  
when expanded in the basis ${\widehat {\sf Q}}^{n}$ dual to the original quorum,
\begin{equation}
{\hat {\rho}} = 
    \frac{1}{2s+1} \sum_{n=1}^{N_s} {\sf P}_{n} {\widehat {\sf Q}}^{n} \, ,
\label{expandrhodual}
\end{equation}
where the coefficients ${\sf P}_{n} = \mbox{ Tr } \left[ \hat \rho\,  {\widehat Q}_{n} \right] \equiv  \bra{{\bf n}_{n}} \hat \rho \ket{{\bf n}_{n}}$ satisfy 
\be
0   \leq    {\sf P}_{n}  \leq 1 \, , 
\qquad 1 \leq n \leq N_s \, .
\label{positive!}
\ee
Each of the $N_s$ numbers ${\sf P}_{n}$ has a value less or equal to one due to the normalization of the density matrix, $\mbox{ Tr }[\,  \hat \rho \, ] = 1 $, and since $\hat \rho$ is a positive operator, the ${\sf P}_{n}$ are {\em non-negative} throughout. This is a unique and essential feature of the basis $\{ {\widehat {\sf Q}}^{n} \}$---the expansion coefficients of $\hat \rho$  with respect to neither the original basis $\{ {\widehat Q}_{n} \}$ nor the multipole bases $\{{\widehat K}_{lm}\}$ or $\{ {\widehat {\sf K}}^{lm}\}$ have this property. The interpretation of the coefficients 
$ {\sf P}_{n}$---to measure the value $s$ along the axis ${\bf n}_{n}$---is clearly compatible with (\ref{positive!}). It is important to note that, although each of the 
 $ {\sf P}_{n}$ {\em is} a probability, they do {\em not} sum up to unity:  
\be
0 < \sum_{n=1}^{N_s} {\sf P}_{n} < (2s+1)^2 \, .
\label{largersum}
\ee
This is due to the fact that they all refer to {\em different orientations} of the Stern-Gerlach apparatus, being thus associated with the measurement of {\em incompatible} observables, 
\be
\left[ {\widehat Q}_{n}, {\widehat Q}_{n'} \right] 
          \neq 0 \, ,
\qquad 1 \leq n,n' \leq N_s \, ,
\label{commutator}
\ee
since the scalar product $\bra{{\bf n}_{n}} {\bf n}_{n'} \rangle$ of two coherent states is different from zero. The sum in (\ref{largersum}) cannot take the value $(2s+1)^2 $ since this would require a common eigenstate of all the operators ${\widehat Q}_{n}$ which does not exist due to 
(\ref{commutator}). By an appropriate choice of the directions ${\bf n}_{n}$ (all in the neighborhood of one single direction ${\bf n}_0$, say), the sum
can be arbitrarily close to $(2s+1)^2 $ for states `peaked' about ${\bf n}_0$. Similarly, the sum of all ${\sf P}_{n}$ cannot take on the value zero since this would require a vanishing density matrix which is impossible \cite{anatole99}. If, however, considered as a sum of {\em expectation values}, there is no need for the numbers ${\sf P}_{n}$ to sum up to unity. Nevertheless, they are not completely independent when arising from a statistical operator: its normalization  implies that 
\be
 \mbox{ Tr } \left[ \, \hat \rho \, \right]  
  = \frac{1}{2s+1} \sum_{n=1}^{N_s} 
        \mbox{ Tr } \left[ {\widehat {\sf Q}}^{n} \right] {\sf P}_{n} 
  = 1 \, ,
\label{restriction}
\end{equation}
which turns one of the probabilities into a function of the $(2s+1)^2-1 = 4s(s+1)$ others, leaving us
with the correct number of free real parameters needed to specify a density matrix. 

It is useful to visualize the description of a density matrix by the numbers ${\sf P}_{n}$ in geometrical terms. Consider the linear space ${\cal A}_s$ of dimension $N_s$, each axis being associated with one projector ${\widehat {\sf Q}}^{n}$ and the  coefficient ${\sf P}_{n}$. Since $\mbox{ Tr } [ {\widehat Q}_{n} {\widehat Q}_{n'} ] = 
| \bra{{\bf n}_{n}} {\bf n}_{n'} \rangle |^2 \neq 0$, this is not an orthonormal basis of ${\cal A}_s$, and neither is its dual $\{ {\widehat {\sf Q}}^{n}\}$. According to (\ref{expandrhodual}) and (\ref{positive!}) each statistical operator determines a point $\vec{\sf P}$ with  components ${\sf P}_{n}$ in an 
$N_s$-dimensional parallelepiped \cite{greub63}. Eq.\ (\ref{restriction}) may be understood as a scalar product of $\vec{\sf P}$ and the vector $\vec{{E}}$ with components ${E}^{n} = \mbox{ Tr } [ {\widehat E} {\widehat {\sf Q}}^{n} ]= \mbox{ Tr } [ {\widehat {\sf Q}}^{n} ] $, where $\widehat E$ denotes the unit operator in ${\cal A}_s$. Thus, the points $\vec{\sf P}$ which correspond to normalized density matrices are necessarily  located in the {\em intersection} $\cal R$ of the parallelepiped with a hyperplane of dimension $4s(s+1)$ in ${\cal A}_s$. However, not all points in $\cal R$ are associated with a density matrix. To see this, imagine the quantum system to be in an eigenstate $ \ket{s, {\bf n}_{n_0}}$ of the projection operator ${\widehat Q}_{n_0}$, say. Then, the corresponding probability ${\sf P}_{n_0}$ has the value one, and all the others are smaller than one. This is the {\em only} point of the unit cube with ${\sf P}_{n_0}=1$ associated with a density matrix but one constructs easily other points satisfying (\ref{restriction}). 
  
Let us turn to the dynamics of the quantum system expressed by the probabilities 
${\sf P}_{n}$. Their time derivative, $d {\sf P}_{n}/ dt = \bra{{\bf n}_{n}}d {\hat \rho} /dt \ket{{\bf n}_{n}}$, is determined unambiguously by Eq.\ (\ref{schroe}). Using the expansion (\ref{expandrhodual}) it is easy to express the resulting equations in the form (\ref{newdynamics}). A {\em closed} set of equations for the variables ${\sf P}_{n}(t)$ follows from plugging (\ref{expandrhodual}) into the right-hand-side of Eq.\ (\ref{schroe}) and taking the expectation value in the state $\ket{ {\bf n}_{n}}$:
\be
\frac{d}{dt} {\sf P}_{n}(t)
        = \frac{i}{\hbar} \sum_{n'=1}^{N_s}  
             \bra{{\bf n}_{n}}
               [  {\widehat {\sf Q}}^{n'}, {\widehat H}]
                         \ket{{\bf n}_{n} } {\sf P}_{n'} (t)\, .
\label{schroeforpro!}
\ee
Thus, the spin dynamics has been expressed entirely in terms of the $N_s$ variables ${\sf P}_{n}$: this equation, the explicit form of (\ref{newdynamics}) for a single spin $s$, provides the main result of this paper. In fact, the dynamics is consistent with (\ref{restriction}): multiply (\ref{schroeforpro!}) by $\mbox{ Tr } [ {\widehat {\sf Q}}^{n} ]$ and sum over all values of $n$:
\be
  \sum_{n=1}^{N_s} 
        \mbox{ Tr } \left[ {\widehat {\sf Q}}^{n} \right] \frac{d}{dt} {\sf P}_{n} 
= \sum_{n'=1}^{N_s}  \mbox{ Tr } \left[  {\widehat {\sf Q}}^{n'}, {\widehat H} \right]
                         {\sf P}_{n'}        
= \frac{i}{\hbar} \mbox{ Tr } \left[ [ \hat \rho , \widehat H ] \right] = 0 \, ,
\label{normcons}
\ee
using (\ref{schroeforpro!}) and expanding the identity as ${\widehat E} =$ $\sum_{n} \mbox{ Tr } [ \widehat{\sf Q}^{n} ] {\widehat Q}_{n}$. Consequently, the time evolution of the quantum system can be represented by a point moving 
in the domain $\cal R$, with a trajectory determined by (\ref{schroeforpro!}). Eq.\ (\ref{schroeforpro!}) will be called the `expectation-value representation' of the equation of motion (\ref{schroe}).

Let us point out some properties of the time evolution of the spin $s$ when given in the expectation-value representation. The dynamical law (\ref{schroeforpro!}) is a {\em closed} set of {\em linear} equations for the $N_s$ real variables ${\sf P}_{n}$: the time derivatives $d {\sf P}_{n}/dt$ at time $t$ are expressed entirely in terms of the probabilities ${\sf P}_{n}$ at that time. Introduce a real $N_s\times N_s$ matrix $\sf M$ with entries 
\be
{{\sf M}_{n}}^{n'} 
   = \frac{i}{\hbar} 
          \bra{{\bf n}_{n}} [ {\widehat {\sf Q}}^{n'}, {\widehat H} \, ] \ket{{\bf n}_{n}} 
         = \frac{i}{\hbar} 
             \mbox{ Tr } \left[{\widehat H} [ {Q}_{n}, {\widehat {\sf Q}}^{n'} ]
                        \right] 
=  \left( {{\sf M}_{n}}^{n'}\right)^* \, ,
\label{niceM}
\ee
using the cyclic property of the trace. Then, one can rewrite the dynamics (\ref{schroeforpro!}) and the constraint (\ref{restriction}) as 
\be
\frac{d\vec{\sf P}(t)}{dt} = {\sf M} \, \vec{\sf P} (t) \, , \qquad 
                          \vec{\sf P} (t_0) \cdot \vec{E} =1 \, .
\label{dynamicalsystem}
\ee
Therefore, the quantum dynamics of a spin $s$ is equivalent to that of a {\em classical dynamical system} with $N_s$ degrees of freedom, constrained to move in a certain region $\cal R$ to be considered as its phase space. For an eigenstate of the Hamiltonian $\widehat H$ with eigenvalue $\epsilon_k$ and density matrix $\hat \rho^{(k)} = \ket{\epsilon_k} \bra{\epsilon_k}$, one has $[\hat \rho^{(k)}, \widehat H] = 0$; hence, the flow generated by ${\sf M}$ in ${\cal R}$ has precisely  $(2s+1)$ `fixed points' with coordinates  ${\sf P}_{n}^{(k)}$. A more detailed study of the flow in (\ref{dynamicalsystem}) will exploit the existence of a metric in the space ${\cal A}_s$, induced by the transformation between the original basis and its dual: ${\widehat Q}_n = 
(2s+1)^{-1} \sum_{n'} {\sf G}_{nn'} {\widehat {\sf Q}}^{n'}$. This metric, ${\sf G}_{nn'} = \mbox{ Tr } [ {\widehat Q}_{n} {\widehat Q}_{n'} ]$, can be shown to define a {\em positive definite} quadratic form.

Somewhat surprisingly, the function $\cal D$ introduced in (\ref{newdynamics}) is {\em linear} in the variables ${\sf P}_{n}$ which, in turn, are 
{\em linear} functions of the density matrix ${\hat \rho}$. Therefore, the {\em convexity} of the state space, ${\hat \rho}^{(\lambda)} = (1-\lambda) {\hat \rho}^{(a)} 
+ \lambda {\hat \rho}^{(b)}$, $0\leq \lambda \leq 1$, turns into: 
\be
\vec{\sf P}^{(\lambda)} 
        = \vec{\sf P}^{(a)} + \lambda  \left( \vec{\sf P}^{(b)} - \vec{\sf P}^{(a)} \right) \, , 
\label{convex}
\ee
tracing out a straight line in the space of expectations.

In the expectation-value representation, the distinction between different pictures of quantum mechanics (Schr\"odinger, interaction, Heisenberg) no longer applies since the quantum dynamics has been expressed entirely in terms of {\em observable} quantities. Furthermore, in this representation there is {\em no} temptation to attach a wave function to an individual quantum spin as is done in the `individual' interpretation of quantum mechanics \cite{jammer89}. From the outset, the involved probabilities make sense only when referring to an (infinite) {\em ensemble} of spin systems prepared identically. Therefore, a vector $\vec{\sf P}$ is associated rather with the procedure of state preparation than with an individual microscopic system. Turned around, the expectation-value representation seems to favour the `statistical interpretation' of quantum mechanics \cite{jammer89}.

Conceptually, the `realization' introduced here differs from other formulations of quantum mechanics `without wave function' such as the phase-space representation through Wigner functions, be it for a particle \cite{baker58} or a spin \cite{wooters87}.
The occurrence of {\em negative} values is characteristic of `quasi-probability' distributions \`a la Wigner, expressing the impossibility that position and momentum simultaneously have definite values. The expectation-value representation allows one to rephrase the quantal dynamics entirely in terms of {\em directly} observable and {\em non-negative} quantities, defined on $(2s+1)^2$ points of the sphere. This construction can be thought of as discretizing the phase space of the classical spin and associating 
probabilities (which, however, do {\em not} sum up to unity) to the individual points. There is a conceptual link to the `probability representation' for both quantum particles and spins \cite{mancini+97,manko+99} which is based on positive smooth distributions on the classical phase space of the underlying system. It provides, however, a highly redundant description, while the expectation-value representation works with nonredundant information only.
   
From a general point of view, Eqs.\ (\ref{expandquorum}) and (\ref{expandquorumdual}) provide the basis for a {\em symbolic calculus} comparable to the Wigner formalism \cite{wigner32,amiet+91} or to the coherent state representation \cite{perelomov86} of quantum mechanics. Once the vector $\vec{\sf P} (t_0) $ associated with a quantum state $\hat \rho (t_0)$ is known, one can extract the time evolution of arbitrary observables $\widehat {\cal A}(t)$ from $\vec{\sf P}(t)$ without ever invoking $\hat \rho (t)$. The details of this calculus based on the expectation-value representation will be developed elsewhere. Furthermore, it is not difficult to generalize the present approach to {\em non-autonomous} quantum systems described by explicitly time dependent Hamiltonian operators ${\widehat H}(t)$.    

In sum, the expectation-value representation of quantum mechanics, as derived from (\ref{expandrhodual}), is equivalent to any other representation. The 
statistical operator $\hat \rho$ of a quantum spin $s$ is represented by a point on a manifold in an $N_s$-dimensional space parameterized by probabilities or, equivalently,  expectation values. Its time evolution, Eqs.\ (\ref{schroeforpro!}) or (\ref{dynamicalsystem}), is governed by a linear and autonomous classical flow. Therefore, one can describe the quantum dynamics as a smooth trajectory in the space of expectation values. With all unobservable elements eliminated from the theory, the expectation-value representation provides an appealing explicit realization of Schr\"odinger's  remark \cite{schroedinger35} to consider the wave function as a ``Katalog der Erwartung.''
\section*{Acknowledgements}
I would like to acknowledge both helpful discussions with J.-P.\ Amiet and financial support by the {\em Schweizerische Nationalfonds}.  
%
%
%

%

\begin{thebibliography}{99}
\bibitem{phases}
         The phases of the states are fixed by the transformation under 
         the anti-unitary time reversal operator $\widehat T$ :  
         $\widehat T \ket{\mu}=(-1)^{s-\mu}\ket{-\mu}$.
\end{thebibliography}
\end{document}